\pdfoutput=1
\documentclass[aps,prl,groupedaddress,superscriptaddress,showpacs,reprint]{revtex4-1}
\usepackage{amsmath}
\usepackage{hyperref}
\usepackage{graphicx}
\usepackage{stackengine}
\usepackage{color}
\usepackage{subfigure}
\def\bra#1{\langle {#1} \rvert}
\def\ket#1{\lvert {#1} \rangle}
\def\braket#1{\langle {#1} \rangle}
\usepackage{helvet}
\usepackage{times}
\begin{document}

	\title{Phase-Programmable Gaussian Boson Sampling Using Stimulated Squeezed Light}
	
	
	\author{Han-Sen Zhong}
	\thanks{These authors contributed equally to this work.}
	\author{Yu-Hao Deng}
	\thanks{These authors contributed equally to this work.}
	\author{Jian Qin}
	\thanks{These authors contributed equally to this work.}
	\author{Hui Wang}
	\author{Ming-Cheng Chen}
	\author{Li-Chao Peng}
	\author{Yi-Han Luo}
	\author{Dian Wu}
	\author{Si-Qiu Gong}
	\author{Hao Su}
	\author{Yi Hu}
	\affiliation{Hefei National Laboratory for Physical Sciences at Microscale and Department of Modern Physics, University of Science and Technology of China, Hefei, Anhui, 230026, China}
	\affiliation{CAS Centre for Excellence and Synergetic Innovation Centre in Quantum Information and Quantum Physics, University of Science and Technology of China, Hefei, Anhui, 230026, China}

	\author{Peng Hu}
	\author{Xiao-Yan Yang}
	\author{Wei-Jun Zhang}
	\author{Hao Li}
	
	\affiliation{State Key Laboratory of Functional Materials for Informatics, Shanghai Institute of Micro system and Information Technology (SIMIT), Chinese Academy of Sciences, 865 Changning Road, Shanghai, 200050, China}
	
	\author{Yuxuan Li}
	\affiliation{Department of Computer Science and Technology and Beijing National Research Center for Information Science and Technology, Tsinghua University, Beijing, China}
	
	\author{Xiao Jiang}

	\affiliation{Hefei National Laboratory for Physical Sciences at Microscale and Department of Modern Physics, University of Science and Technology of China, Hefei, Anhui, 230026, China}
	\affiliation{CAS Centre for Excellence and Synergetic Innovation Centre in Quantum Information and Quantum Physics, University of Science and Technology of China, Hefei, Anhui, 230026, China}

	\author{Lin Gan}
	\author{Guangwen Yang}
	\affiliation{Department of Computer Science and Technology and Beijing National Research Center for Information Science and Technology, Tsinghua University, Beijing, China}

	\author{Lixing You}
	\author{Zhen Wang}
	
	\affiliation{State Key Laboratory of Functional Materials for Informatics, Shanghai Institute of Micro system and Information Technology (SIMIT), Chinese Academy of Sciences, 865 Changning Road, Shanghai, 200050, China}
	
	\author{Li Li}
	\author{Nai-Le Liu}
	
	\affiliation{Hefei National Laboratory for Physical Sciences at Microscale and Department of Modern Physics, University of Science and Technology of China, Hefei, Anhui, 230026, China}
	\affiliation{CAS Centre for Excellence and Synergetic Innovation Centre in Quantum Information and Quantum Physics, University of Science and Technology of China, Hefei, Anhui, 230026, China}
	
	\author{Jelmer J. Renema}
	
	\affiliation{Adaptive Quantum Optics Group, Mesa+ Institute for Nanotechnology, University of Twente, P.O. Box 217, 7500 AE Enschede, Netherlands}
	
	\author{Chao-Yang Lu}
	\author{Jian-Wei Pan}
	
	\affiliation{Hefei National Laboratory for Physical Sciences at Microscale and Department of Modern Physics, University of Science and Technology of China, Hefei, Anhui, 230026, China}
	\affiliation{CAS Centre for Excellence and Synergetic Innovation Centre in Quantum Information and Quantum Physics, University of Science and Technology of China, Hefei, Anhui, 230026, China}
	
	\date{\today}
	
	\begin{abstract}
We report Gaussian boson sampling (GBS) experiments which produce up to 113 photon detection events out of a 144-mode photonic circuit. A new high-brightness and scalable quantum light source is developed, exploring the idea of stimulated emission of squeezed photons, which has simultaneously near-unity purity and efficiency. This GBS is programmable by tuning the phase of the input squeezed states. We demonstrate a new method to efficiently validate the samples by inferring from computationally friendly subsystems, which rules out hypotheses including distinguishable photons and thermal states. We show that our noisy GBS experiment passes a nonclassicality test using an inequality, and we reveal non-trivial genuine high-order correlation in the GBS samples, which are evidence of robustness against possible classical simulation schemes. The photonic quantum computer, \textit{Jiuzhang} 2.0, yields a Hilbert space dimension up to $\sim 10^{43}$, and a sampling rate $\sim 10^{24}$ faster than using brute-force simulation on supercomputers.
\end{abstract}
	
	
	\maketitle

\begin{figure*}[!htp]
    \centering
    \includegraphics[width=0.9\linewidth]{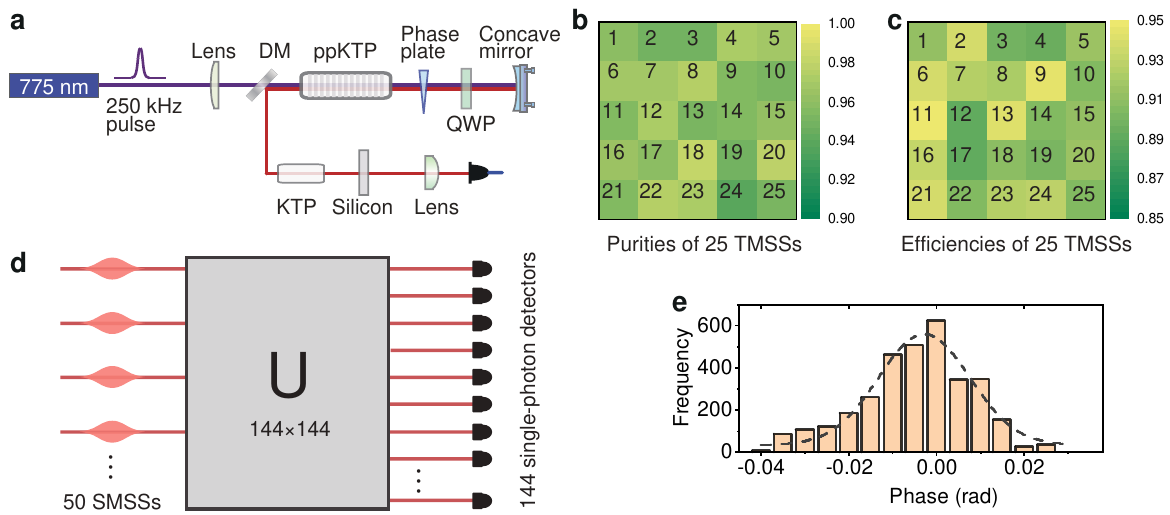}
    \caption{The key experimental parameters. (a) Squeezed light amplification by stimulated parametric down-conversion. Both the pump and parametric photons are refocused by the concave mirror back to the PPKTP for stimulated emission. The down-converted photons are polarization-exchanged by the QWP and its relative phase with pump light are tuned by the wedge phase plate. The squeezed light passed a KTP crystal for birefringence compensation and an antireflection coated silicon plate for filtering out residual pumping light, before being collected into the single-mode fiber. (b) The purities of the 25 squeezed light sources with an average of 0.961. (c) The collection efficiencies of the 25 squeezed light sources with an average of 0.918. (d) An overview of the GBS set-up. 25 pairs of two-mode squeezed photons are sent into a 144-mode interferometer, and the output distribution is readout by 144 single-photon detectors. (e) Phase stability of the whole set-up. The histogram shows that the phase fluctuations are within $\lambda/100$ in an hour.
    }
    \label{fig:1}
\end{figure*}

	The tantalizing promise of quantum computational speedup in solving certain problems has been strongly supported by recent experimental evidence from a high-fidelity 53-qubit superconducting processor  \cite{arute2019quantum} and Gaussian boson sampling \cite{zhong2020quantum,aaronson2011computational,hamilton2017gaussian} (GBS) with up to 76 detected photons. Analogous to the increasingly sophisticated Bell tests \cite{bell1964einstein} that continued to refute local hidden variable theories \cite{einstein1935can}, quantum computational advantage tests \cite{preskill2012quantum,harrow2017quantum,lund2017quantum} are expected to provide increasingly compelling experimental evidence \cite{arute2019quantum,zhong2020quantum} against the Extended Church-Turing thesis \cite{bernstein1993quantum}. In this direction, continued competition between upgraded quantum hardware and improved classical simulations \cite{pednault2019leveraging,clifford2020faster,huang2020classical,quesada2020quadratic,zhou2020limits,renema2020marginal,pan2021simulating,gray2021hyper} are required.
	
Boson sampling, proposed by Aaronson and Arkhipov \cite{aaronson2011computational}, is an intermediate model of linear optical quantum computation \cite{kok2007linear,knill2001scheme}. Realizing boson sampling with a level of post-classical computational complexity requires high-performance quantum light sources, a large-scale, low-loss photonic circuit, and high-efficiency single-photon detectors, all of which are essential building blocks for universal quantum computation using photons. Gaussian boson sampling (GBS) exploits squeezed vacuum states as input non-classical light sources, with a significant advantage of dramatically increasing the output multi-photon click probability \cite{hamilton2017gaussian,lund2014boson}. Experimentally, generating an increasingly large array of squeezed states with near-unity photon indistinguishability and collection efficiency, and sufficiently high brightness, simultaneously, is still a non-trivial challenge \cite{zhong2020quantum,zhong2019experimental,arrazola2021quantum,paesani2019generation}. To increase the number of input squeezers or their brightness, one typically uses stronger pump laser power, or, if the total power is fixed, narrows the focus waist. However, the stronger pump power and tighter focus could result in self-focusing and self-phase modulation that lower both the photon purity and the collection efficiency. Due to this problem, in the previous GBS experiment \cite{zhong2020quantum}, band-pass filters were used to increase the photon indistinguishability to 0.938 at the cost of decreasing the collection efficiency to 0.628.

\textit{Stimulated emission of squeezed photons.}---To overcome this limitation, inspired by the concept of light amplification by stimulated emission of radiation (LASER), we design a new, scalable quantum light source based on stimulated emission of squeezed states. A schematic drawing is displayed in Fig. 1a. The idea is that spontaneously generated photon pairs, in resonance with the pump laser, stimulate the parametric emission of the second photon pair in a gain medium \cite{lamas2001stimulated}. In our experiment, transform-limited laser pulses at a wavelength of ~775 nm are focused on PPKTP crystals to generate two-mode squeezed states (TMSS). After the PPKTP, the pump laser and the collinear TMSS photons are reflected back by a concave mirror, which are used as seeds to stimulate the second parametric process. The birefringence walk-off between the horizontally and vertically polarization photons of the TMSS is compensated using a quarter-wave plate (see Fig. S1). The dispersion between the pump laser and the TMSS is compensated by eliminating the frequency correlation in the design of the PPKTP crystals (Fig. S2).

By tuning a quartz wedge plate to change the relative phase between the reflected pump laser and the TMSS, we observe an interference fringe of the brightness of the TMSS (Fig. S3). The interference visibility is 0.951, showing the degree of mode matching of the seed and the stimulated TMSS. At constructive interference, the stimulated process increases the brightness of the source by a factor of $\sim 3$ compared to the previous single-pass scheme. Equivalently, we can lower the pump power and use large focus waist to generate TMSS with both higher indistinguishability and collection efficiency. In this work, at a waist of 125 $\mu$m (65 $\mu$m), the measured collection efficiency is 0.918 (0.864), and the photon purity is 0.961 (0.946) simultaneously, without any narrowband filtering (see Fig. 1b,c). We note that our double-pass scheme can be straightforwardly extended to higher orders and generate higher brightness, which can serve as a scalable and near-optimal quantum light source.

\textit{144-mode phase-locked interferometer.}---We generated 25 TMSS sources using the stimulated scheme (see Fig. S4 for their phase and squeezing parameters), and send them into a 144-mode interferometer to implement random unitary transformations (see Fig. 1d and Fig. S5 for an illustration of the whole set-up). The compact 3-dimensional interferometer \cite{wang2019boson} includes 2 polarization modes and 72 spatial modes. The interferometer features a very high transmission rate of 96.5\%. Taking into account of the single-mode fiber coupling, the additional polarizing beam splitter, wave plates, and filters, the overall efficiency is 78\%. The relative path length difference between different ports is calibrated within a standard deviation of 1.6 $\mu$m (Fig. S6), indicating that the wave packet overlap inside the interferometer is better than 0.999. After the interferometer, the output photons are detected by 144 superconducting nanowire single-photon detectors with an average efficiency of 83\% (Fig. S6).

In contrast to Fock-state boson sampling \cite{aaronson2011computational} where there is no phase relation between single photons, GBS relies on coherent superposition of the photon numbers. To this aim, we develop active phase locking over the whole optical path and passive stabilization inside the interferometer. For the active locking, as displayed in Fig. S7, a continuous-wave 1450-nm laser is split into 26 paths; one is used as a reference, and the other 25 are combined and co-propagates with the 25 TMSSs. We implement an optical phase-locking loop  \cite{suppl} for each TMSS. The phase fluctuation of the whole system is controlled within $\lambda/100$ for a duration of one hour, as shown in Fig. 1e. Such a phase instability will cause a $\sim 0.5\%$ decrease of the photon interference visibility.

Fig. S8 plots the reconstructed matrix of the $144 \times 144$ interferometer. To confirm if the obtained matrix is unitary, we calculate the product with its Hermitian conjugate, which gives an identity matrix (Fig. S9). We further compare the Fig. S8 with the ideal Haar-random matrix elements, which shows a good agreement (see Fig. S10).

\begin{figure*}[!htp]
    \centering
    \includegraphics[width=1\linewidth]{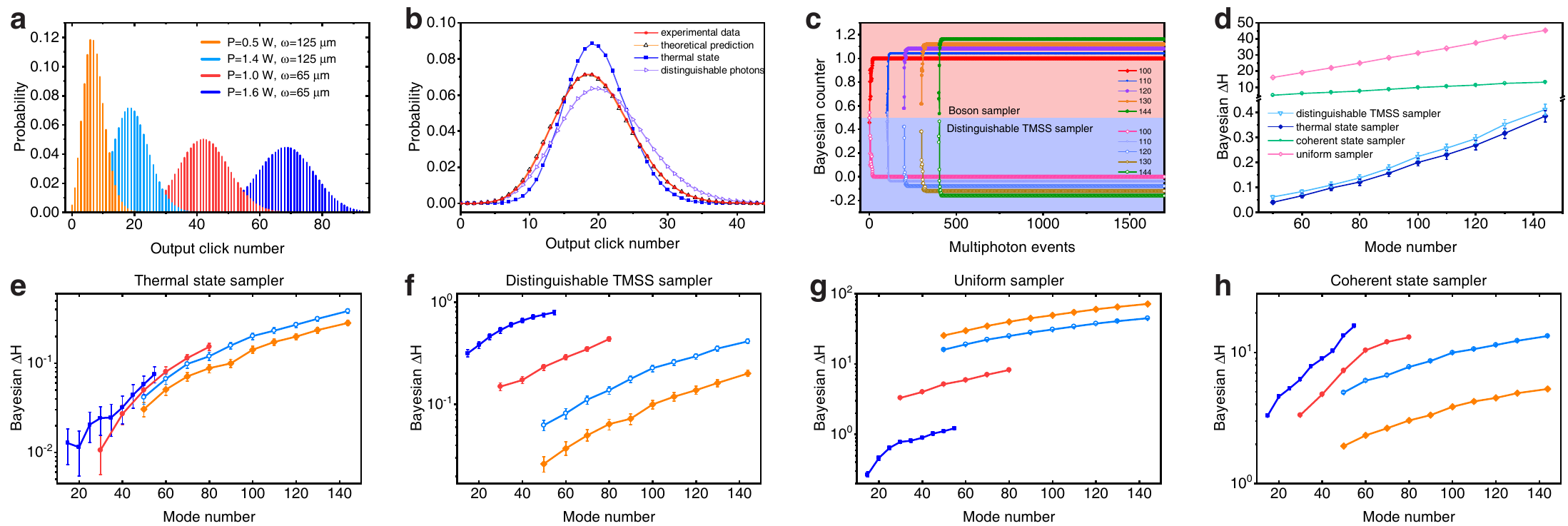}
    \caption{Validation of GBS. (a) Photon clicks distribution at different pump power and focal waists. (b) Output photon-click distribution of the experiment (red), theoretical prediction (orange), thermal state (blue), and distinguishable photons (purple). The experimental GBS data perfectly overlaps with the theoretical curve, while strongly deviates from the other three mockups. (c) Bayesian test against the hypothesis of distinguishable photons with subsystem mode number ranging from 100 to 144. (d) The calculated Bayesian $\Delta H$ as a function of the subsystem mode number. For all the hypotheses, it is clear that the validation confidence is stronger for a larger mode number. (e-h) The experimental results of Bayesian $\Delta H$ at different laser pump power and focal waists for the validations against the thermal state (e), distinguishable photons (f), uniform samplers (g), and coherent state (h). The color coding is the same as panel (a). Error bars represent one standard deviation of uncertainty.
    }
    \label{fig:2}
\end{figure*}

\textit{Validating GBS.}---We perform GBS at different laser pump power (from 0.15 W to 1.6 W) and focal waists (65 $\mu$m and 125 $\mu$m), which gives rise to different photon number distributions. Four examples are shown in Fig. 2a, where the maximal output click number ranges from 32 to 113.

To validate the obtained samples, we hope to further provide evidence for the correct operation of the GBS, and rule out mockups using possible spoofing methods \cite{aaronson2013bosonsampling,bentivegna2015bayesian,spagnolo_experimental_2014}. The most plausible hypotheses in this experiment include using coherent light sources (lasers) as input, distinguishable photons (owing to mode mismatch and imperfect light sources), thermal states (due to excessive photon loss), and uniform random outcome.

First, we compare the total output click number distribution with the possible mockups. Fig. 2b shows that the experimental GBS data perfectly overlaps with the theoretical prediction, while strongly deviates—in both their line shapes and peak positions—from the classical mockups based on distinguishable photons and thermal states.

Having studied the ensemble distributions, we proceed to validate the individual multi-photon click samples. The validation protocols, such as Bayesian test \cite{bentivegna2015bayesian}, likelihood test \cite{spagnolo_experimental_2014}, and heavy output generation \cite{zhong2020quantum}, require the calculation of the probability of each sample, that is, the corresponding submatrix Torontonian. Such calculations are exponentially hard for classical computation, and the threshold is around 40 photons. However, our experiment involves much larger number of photon clicks, reaching a maximal of 113. To tame the validation in the intractable regime, we propose a computationally friendly method. The idea is that, we start from a subsystem with fewer output bosonic modes, which is computationally easier. Intuitively, the strength of the validation will become stronger if there are less modes traced out. Therefore, if one can rule out mockups even in the subsystem and observe an increasing trend when the mode number increase, then the confidence of the whole system—although cannot be computed directly—will be stronger.

In the standard Bayesian test \cite{bentivegna2015bayesian}, for each measured event k, we use $Q_k$ and $R_k$ to denote the probability associated with the GBS and a mockup sampler. We define Bayesian counter $C_B$ as:
 \begin{equation}
 C_B = \chi_N/(\chi_N + 1),\text{ where~} \chi_N = \prod_{k=1}^N \frac{Q_k}{R_k}.
 \end{equation}
$\chi_N>1$  indicates that the experimental samples are more likely from the GBS than the mockup. $C_B$ is the probability that the samples are from the GBS after testing $N$ events. We further define Bayesian: $\Delta H = \log \chi_N/N$, to measure the strength of the validation for unit samples. A larger value of $\Delta H$  indicates a larger deviation between the GBS and the mockups \cite{google_simulation_unpublished}.
  
Fig. 2c shows a typical example of the Bayesian test for subsystem size from 100 to 144, which is used for the validation of the sample against mockups using distinguishable photons. The Bayesian counter $C_B$ grows rapidly and reaches over 99.6\% after 50-100 events. The Bayesian tests against other hypotheses are plotted in Fig. S11.

\begin{figure*}[!htp]
    \centering
    \includegraphics[width=0.9\linewidth]{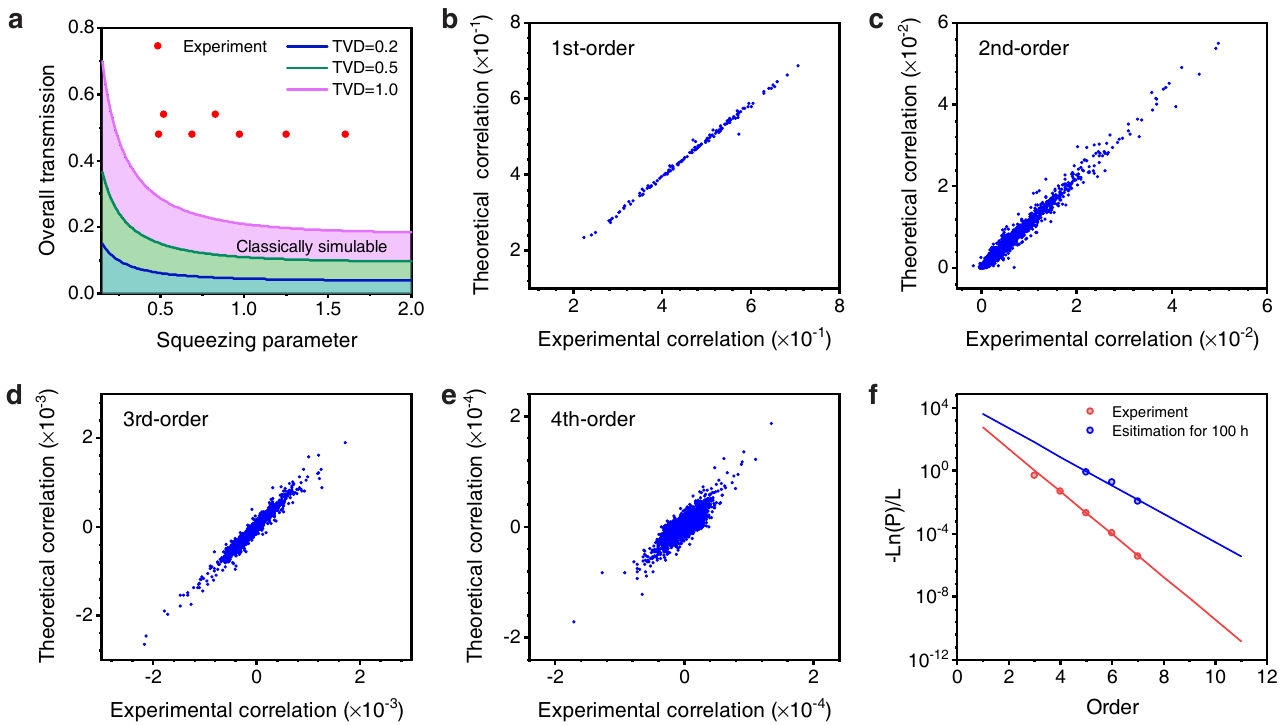}
    \caption{Truncated correlation functions of the GBS machine. (a) Nonclassicality test of experiment. The colored areas correspond to the classically simulable regimes, while our experiment results (red dots) are well above color area. (b)-(e) are the truncated first- to fourth-order correlation functions of calculated theoretical values ($y$ axis) and extracted experimental results ($x$ axis), respectively. The results clearly indicate that our data have genuine correlations. (f) Extracted p-values of Spearman's rank-order tests. The vertical axis is $- \ln (p)/L$, where $L$ is the number of samples. The red dots are from our experimental dataset within 200 s, which drops exponentially when the correlation order linearly increase. By fitting, we extrapolate that there is up to correlations of $19 \pm 1$ order in our experiment for $p<0.05$. The blue dots are from a hypothetic dataset within 100 hours. Using the same method, we estimate that correlations of $43\pm7$ order can be determined.
    }
    \label{fig:3}
\end{figure*}

The low pumping intensity regime with less than 40 maximal photon clicks is within the computational capability of classical supercomputers. Thus, we can validate the data over the full range of photon clicks and mode number. Fig. 2d shows the calculated Bayesian $\Delta H$ as a function of the subsystem mode size from 50 to 144, which, for all the hypotheses, clearly shows an increasing validation confidence for larger mode number.

Next, we move to the computationally intractable regime under high pumping intensity. The Bayesian $\Delta H$ are calculated and plotted with varying mode size for the validations against the thermal states (Fig. 2e), distinguishable photons (Fig. 2f), uniform samplers (Fig. 2g) and coherent state (Fig. 2h). At the high laser intensity regime, the supercomputer can only handle the data at the subspaces with 20-80 output bosonic modes. For all the tested data, we observe that not only $\Delta H>0$, but also it follows the same increasing trend as in the low intensity regime. This allows us to infer that the full mode samples in the quantum advantage regime that share the same set-up can be validated with a stronger confidence.

\textit{Limits on classical simulability of GBS.}---GBS is susceptible to experimental imperfections which might degrade the quality of the photon interference to the point that the experiment becomes classically simulable \cite{qi2020regimes,garcia2019simulating,renema2020simulability}, in the sense that a classical efficient strategy exists to produce samples which cannot be distinguished to within some chosen accuracy from those coming from the quantum sampler. The main sources of noise in GBS are photon loss, photon distinguishability, noise on the interferometer, and detector dark counts.

For GBS, two main strategies for classical simulation exist. The first uses the non-negativity of quasi-probability distributions (QPD) (generalizations of the Wigner function) as a strategy for simulation. The second uses the fact that in GBS, the marginal distributions of photon numbers (i.e., the probabilities to observe some subset of detection events irrespective of the others) are informative about the complete probability distribution.

For the QDP based simulations, an inequality exists that that demarcates the regime of simulability \cite{qi2020regimes}. Thus, any finite-sized experiment must pass this inequality to show that it is not simulable by this strategy. The inequality is given by \cite{qi2020regimes}:
\begin{equation}
\text{sech} \left\{  \frac{1}{2} \Theta \left[ \ln\left( \frac{1-2 q_D}{\eta \text{e}^{-2r}+1-\eta} \right)  \right]  \right\} > \text{e}^{-\epsilon^2/4K},
\end{equation}
where $r$ is the squeezing parameters, $\eta$ is the overall photon transmission rate, $K$ is the number of squeezed sources, $\epsilon$ is the total variance distance of the experimental GBS samples compared to the ideal cases, $\Theta$ is the ramp function $\Theta (x) :=\max(x,0)$, and  $q_D=p_D / \eta_D$ is calculated from the photon detector efficiency ($\eta_D$) and dark count rate ($p_D$). Summarizing the efficiencies at the quantum light sources, interferometer, and detectors, the overall transmission rate in our experiment is 48\% and 54\% at the focal waists of 65 $\mu$m and 125 $\mu$m, respectively. As we plot in Fig. 3a, this inequality is violated for any $\epsilon$ for all the chosen parameters, which makes our experiment pass the nonclassicality test.

\begin{figure*}[!htp]
    \centering
    \includegraphics[width=0.9\linewidth]{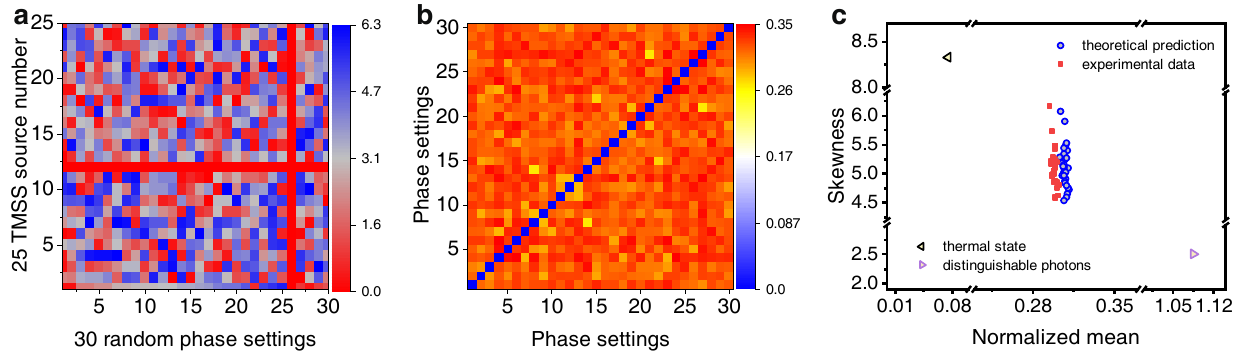}
    \caption{The programmability of the GBS machine. (a) 30 different groups of phase settings of all 25 TMSSs. All phase curves are stacked with a $2\pi$ offset in the vertical axis direction. (b) Total variance distances between experimental two-point correlation functions with different groups of TMSS phases. The average of diagonal elements is 0.068(1), which represents the statistical fluctuating noise due to finite samples. The average of off-diagonal elements is 0.32(1), represent successfully distinguishing samples with 30 groups of TMSS phases. (c) Normalized mean and skewness of the distributions of two-point correlation functions, including experimental samples (red square), theoretical predictions (blue dot), thermal state (black triangle) mockup and distinguishable photons (purple triangle) mockup. Each point is from one setting of the TMSS phases.
    }
    \label{fig:4}
\end{figure*}

The second family of simulation strategies  \cite{renema2020marginal,renema2020simulability,popova2021cracking,google_simulation_unpublished} relies on low-order information of the probabilities to estimate the true distribution, such as using the truncation of Fourier transformation of the interfering probability amplitudes to obtain a polynomial simulation scheme and the marginal distributions of photons which can be computed efficiently  \cite{note_simulation}. The physical picture in the GBS is that interference processes of $n$ photons are represented in the distribution by the $n$-th-order marginal distributions. The computational cost of computing a $k$-th order marginal probability is the same as computing the probability of a $k$-photon boson sampler, i.e. $2^k$, up to a polynomial factor. Since the marginals provide some information about the distribution, it may be possible to construct mockup distributions which resemble the true quantum distribution. For Fock-state \cite{aaronson2011computational} and superposition state boson sampling \cite{renema2020simulability}, it is known how to do this efficiently. For GBS, a straightforward extension of this scheme has so far failed \cite{renema2020simulability}.

Since there is no concrete marginal-based simulation strategy based on GBS, there is no clear equation to validate against. To argue the degree to which our results are robust against simulation schemes based on marginal distributions, we investigate the truncated high-order correlations present in our dataset. The truncated $k$-order correlation quantifies the degree to which the corresponding $k$-body marginal cannot be explained by the marginal output distribution of 1-body to $(k-1)$-body coincidence. Truncated  $k$-order correlation can be recursively defined as \cite{phillips2019benchmarking,duneau1973decrease,Walschaers2018}
\begin{equation}
\begin{aligned}
\kappa&\left(X_1 X_2 \dots X_N\right) \\
&= E\left(X_1 X_2 \dots X_N\right)-
\sum_{p\in P_n} \prod_{b \in p} \kappa\left[\left(X_i\right)_{i\in b}\right],
\end{aligned}
\end{equation}
where $P_n$ represents all the partitions of $\left\{ 1,2,\dots,k \right\} $, except the universal set. Examples of the experimental and theoretical truncated 1-order to 4-order correlations are plotted in Fig. 3b to 3e. It is evident that the experimental results are highly consistent with the theoretical results, indicating the presence of nontrivial genuine high-order correlation in the GBS data. In order to quantitatively extract relatively small high-order correlations from the statistical noise, we use Spearman's rank-order test to calculate $p$-values to against the hypothesis that the experimental and theoretical results are irrelevant. We extrapolated the $p$-value of Spearman's rank-order test (Fig. 3f), and estimated that there are up to $(19\pm 1)$-order correlations for $p<0.05$ already in the data collected within 200 s \cite{reexamine_jiuzhang1}. We estimate that a 100-hour data collection—which is feasible in our set-up—can further reveal up to $(43\pm 7)$-order correlations.

Finally, we comment on whether this level of higher-order correlations is sufficient to protect against classical simulation. If a scheme exists that uses the $k$-th order marginals in an efficient way (e.g., by only computing those marginals associated with a given configuration of outcomes), such a scheme will be able to efficiently simulate our system. For a scheme that requires a full enumeration of all marginals \cite{google_simulation_unpublished}, and then an inversion of the full marginals problem, there is an additional efficiency overhead, since there are combinatorially many such marginals. We estimate that for a classical computer, computing all marginals up to $\sim 10^\text{th}$ order is already unfeasible. Since our experimental data contains nontrivial correlations up to $\sim 19^\text{th}$ order, we expect our results to be robust against such a simulation scheme. We hope that our work will inspire new efforts for quantitative characterizations for the GBS and new classically efficient spoofing methods \cite{qi2020regimes,garcia2019simulating,renema2020simulability,popova2021cracking} to challenge the GBS device. All raw data has been archived online to encourage the development and testing \cite{online}.

\begin{figure}[htbp]
    \includegraphics[width=0.8\linewidth]{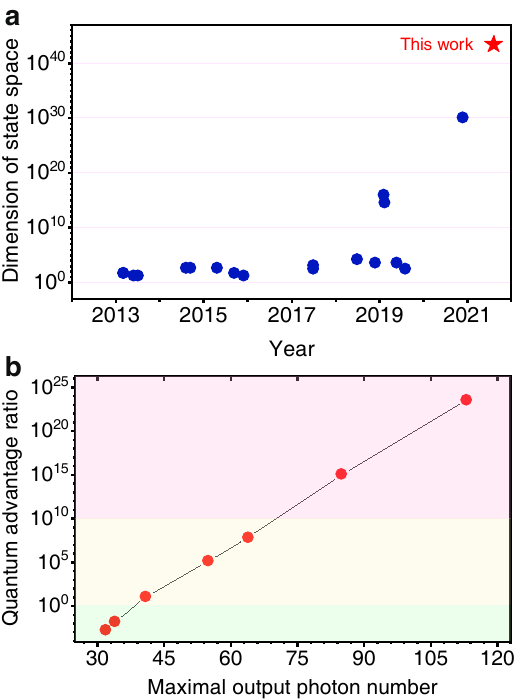}
    \caption{ Dimension of Hilbert space and quantum advantage ratio. (a) Summary of the output Hilbert spaces of boson sampling and random circuit sampling. This work (red star) has a Hilbert space of $\sim 10^{43}$. (b) The quantum advantage ratio (compared to direct simulation with classical computers) as a function of the maximally detected photon clicks. Based on brute-force simulation, a tentative quantum advantage ratio of $10^{24}$ is observed in this work.
    }
    \label{fig:5}
\end{figure}

\textit{Phase-programmable GBS.}---The transformation matrix in the GBS is determined by both the interferometer and the squeezing parameters and phases of the input TMSS. Therefore, by changing the input parameters, the underlying matrix can be reconfigured and the GBS machine can be programmed to solve different parameters. Here, we demonstrate the programmability of the GBS quantum computer by setting 30 different groups of random phases of the input TMSS, as shown in Fig. 4a. To change the phases, we add adjustable electric delay lines to the reference signal to which each TMSS is phase locked.

Due to the huge Hilbert space dimension, it is not feasible to directly compare the 30 groups of output samples with their theoretical distribution. To extract distinguishable statistical properties of the 30 groups of samples, we use two-point correlation function \cite{phillips2019benchmarking}, which is defined as  $C_{i,j} = \left< \Pi_1^i \Pi_1^j\right>-\left< \Pi_1^i \right>\left< \Pi_1^j \right>$, where $\left< \Pi_1^i \right> = \textbf{I}-\ket{0}_i\bra{0}_i$  represent a click in mode $i$. The two-point correlation method is suitable for characterization as it can eliminate the influence of unbalanced amplitudes between different modes, and allows to extract the photon interference terms only.

There are $144\times 143/2=10296$ combinations of  $C_{i,j}$ for each phase setting. To quantify the distance between the 30 groups, we calculate the total variance distances of the $C_{i,j}$  between any two groups, and plot the results in Fig. 4b. The diagonal elements—the settings with the same phase—have an average of 0.068(1), which is the noise level of statistical fluctuations due to finite samples. However, the average of the other elements has an average of 0.32(1), which is significantly larger than the statistical fluctuation, thus successfully distinguishing the data of the 30 different groups.

We further investigate the statistical properties of the two-point correlation functions of the experimental samples, their theoretical predictions, and three mockups with thermal state and distinguishable photons. As shown in Fig. 4c, the horizontal axis is normalized mean in the form of 
$\braket{C} M^2/N^2$, and the vertical axis is skewness written as 
$\left(  \braket{C^3}-3\braket{C^2}\braket{C}+2\braket{C}^3 \right)/ \sqrt{\left( \braket{C^2} - \braket{C}^2 \right)^3   } $,
where $C$ represent the two-point correlation functions, and $M, N$ represent the number of output and input modes, respectively. Each point is from one setting of the TMSS phase. The results are close to the theoretical calculation and clearly far away from the two mockups. Note that the statistical properties of the GBS samples vary with different phases thus have scattered plots, while those of the mockups with thermal state and distinguishable photons are not affected by the phase thus have only one data point.

\textit{The classical computational cost.}---Quantum computing experiments are rapidly moving into new realms of increasing size and complexity. One characteristic is output Hilbert space dimension. Fig. 5a plots the computational state-space dimension of the boson sampling \cite{zhong2020quantum,zhong2019experimental,paesani2019generation,wang2019boson,spagnolo_experimental_2014,broome2013photonic,spring2013boson,tillmann2013experimental,crespi2013integrated,carolan2015universal,wang2017high,loredo2017boson,bentivegna2015experimental,zhong201812,wang2018toward,carolan2014experimental,giordani2018experimental} and random circuit sampling experiments, where the current work reaches a new record to $\sim 10^{43}$.

Finally, we estimate the classical computational cost to simulate the GBS. We choose the benchmarking algorithm \cite{quesada2018gaussian} by calculating Torontonian on the Sunway TaihuLight supercomputer \cite{li2020benchmarking}. For each output multi-photon sample, we calculate the corresponding time cost used by the supercomputer to obtain one sample by calculating the submatrix Torontonians, which is plotted in Fig. S14 for different pumping intensity. The overall classical time cost is then compared to the sample collection time (200 s) using the GBS quantum computer, which we call quantum advantage ratio here. The advantage ratio is summarized in Fig. 5b as a function of the maximally detected photon clicks in the quantum experiments. We observe a transition from no quantum advantage (the ratio is $10^{-3}-10^0$, light green area), modest speedup (the ratio is $10^0-10^{10}$, light yellow area), to overwhelming speedup (the ratio is $10^{10}-10^{24}$, light red area).

We emphasize again that this estimation is based on the direct simulation algorithm \cite{quesada2018gaussian}. With the ongoing development of more efficient classical algorithms \cite{renema2020marginal,clifford2020faster,quesada2020quadratic,popova2021cracking} and possibly by exploiting the photon loss and partial distinguishability \cite{qi2020regimes,garcia2019simulating,renema2020simulability}, we expect and encourage classical algorithmic improvements to narrow the quantum-classical gap. Meanwhile, inspired by the classical algorithmic improvements and possible spoofing methods, the GBS quantum computer will also continue to be upgraded to compete with the classical simulation. For example, the stimulated squeezed light sources developed here can be straightforwardly scaled to higher orders and larger photon numbers, with near-unity efficiency and indistinguishability simultaneously.

\textit{Outlook.}---The GBS links to several potentially applications such as quantum chemistry \cite{huh2015boson,huh2017vibronic,jahangiri2021quantum}, graph optimization \cite{arrazola2018using,arrazola2018quantum,bradler2021graph}, and quantum machine learning \cite{schuld2019quantum,chabaud2021quantum}. By adjusting only the squeezing parameters and phases, the current set-up can already be used for quantum machine learning \cite{schuld2019quantum}. A natural next step would be to use the GBS quantum computer developed here as a special-purpose photonic platform to investigate whether these algorithms can provide any quantum speedup \cite{preskill2018quantum}. Finally, the quantum optical set-up consisting of squeezed states fed into a linear optical network and followed by photon detection can be used for the creation of different family of entangled states \cite{menicucci2006universal}. One prominent example is the intrinsically fault-tolerant Gottesman-Kitaev-Preskill code \cite{su2019conversion}.

\begin{acknowledgments}
We thank S. Aaronson, A. Arkhipov, S. Boixo, J. Martinis, Gil Kalai, G. Kindler, M. Y. Niu, L. Li, H. Neven, A. S. Popova, J. C. Platt, A.N. Rubtsov, V. N. Smelyanskiy, B. Villalonga, and G. Weinstein for critical comments and inspiring discussions. This work was supported by the National Natural Science Foundation of China, the National Key R\&D Program of China, the Chinese Academy of Sciences, the Anhui Initiative in Quantum Information Technologies, the Science and Technology Commission of Shanghai Municipality, and NWO Veni.

\end{acknowledgments}

%

\newpage
\newpage

\begin{figure*}[!htp]
    \centering
    \includegraphics[angle=90,width=0.53\linewidth]{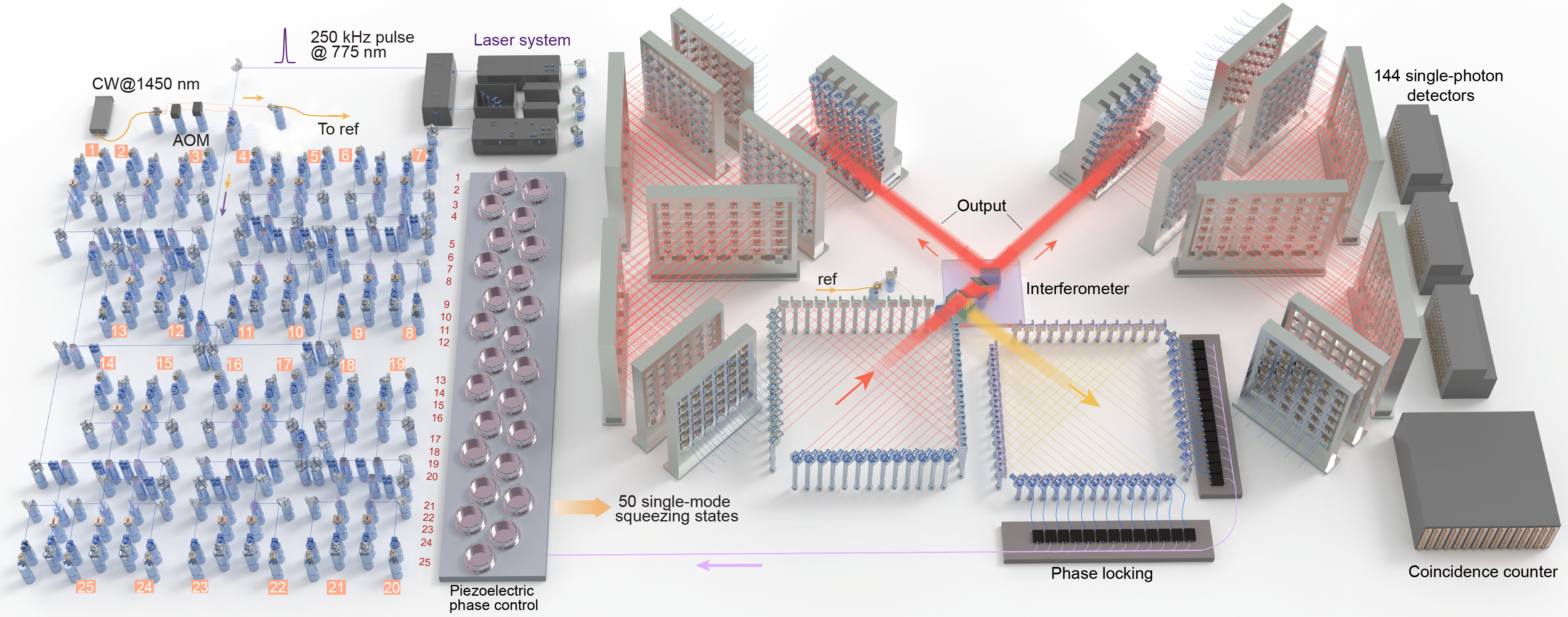}
   
\end{figure*}

\end{document}